# Two-dimensional Decoding Algorithms and Recording Techniques for Bit Patterned Media Feasibility Demonstrations

Yuri Obukhov, Pierre-Olivier Jubert, Daniel Bedau, Michael Grobis

HGST, a Western Digital Company, San Jose, CA 95135

**Recording experiments and decoding algorithms are presented for evaluating the bit-error-rate of state-of-the-art magnetic bit-patterned media. The recording experiments are performed with a static tester and conventional hard-disk-drive heads. As the reader dimensions are larger than the bit dimensions in both the down-track and the cross-track directions, a two-dimensional bit decoding algorithm is required. Two such algorithms are presented in details together with the methodology implemented to accurately retrieve island positions during recording. Using these techniques, a 1.6 Td/in$^2$ magnetic bit pattern media is demonstrated to support 2D bit error rates below 1e-2 under shingled magnetic recording conditions.**

## I. Introduction

**B**IT PATTERNED MEDIA (BPM) is one of the promising technologies for enabling higher areal density (AD) in hard disk drive based magnetic data storage [1][2][3]. The main advantage of BPM over granular media is that it allows better thermal stability and increased signal to noise ratio at equivalent areal density [4][5][6] . The main disadvantages of BPM are the system architecture requirements of write synchronization and the fabrication challenge of creating highly uniform and well-ordered nanoscale islands over macroscopic distances. Recent BPM technological demonstrations at areal densities of 1.5 Td/in2 (Tera-dot per square inch) and above have shown that many of these challenges are surmountable and illustrate the promise of this technology [7][8][9][10].

In this paper we present several experimental and analytical methods developed for the recording evaluation of high density BPM using a static tester in which the read-write head is in contact with the disk [11][12]. The static tester enables characterization of early prototype BPM media, but achieving the sub-nm positioning resolution required for proper recording evaluation is challenging. In addition, the track pitches in high density BPM are narrower than the magnetic read widths (MRW) of the readers found in commercially available hard disk drive (HDD) read/write heads. The large inter-track interference (ITI) during read back makes typical one dimensional (1D) decoding techniques unusable [13]. Overcoming these problems required the development of image processing, two dimensional (2D) decoding, and servo positioning methods, which we present in this paper. We discuss these techniques in the context of a 1.6 Td/in$^2$ recording feasibility demonstration.

### A. Experiment

The 1.6 Td/in$^2$ BPM medium in this work consists of discrete magnetic islands fabricated using nanoimprint lithography. Details of the fabrication process and additional sample information can be found in Ref [10] The islands are arranged in a rectangular lattice with an 18.5-nm pitch down-track and a 22-nm pitch cross-track. The island size and island down-track and cross-track position distributions are Gaussian with standard deviations of 1 nm, 1.1 nm and 1.2 nm, respectively. The sample defect rate (missing and merged islands) is lower than 1e-3. The magnetic material is an alloy of CoCrPt with a fill factor after etching of approximately 65%. The medium contains a soft magnetic underlayer (SUL) below the magnetic islands.

The recording measurements were performed using a static tester that allows accurate positioning of the read/write elements of a commercially available HDD head that is in contact with the disk. The scan speeds are ~100 µm/s. The spacing between the surface of the read/write elements and the top of the magnetic layer is difficult to measure, but is estimated to be around 10 nm. The bits are written by applying a current to the writer coil, with a typical pulse duration of 50 ns. The written track width spans several bits in the cross track direction. The reader magnetic read width in the cross track direction is ~60% larger than the track pitch. The down track magnetic read width is also ~60% larger than the bit pitch.

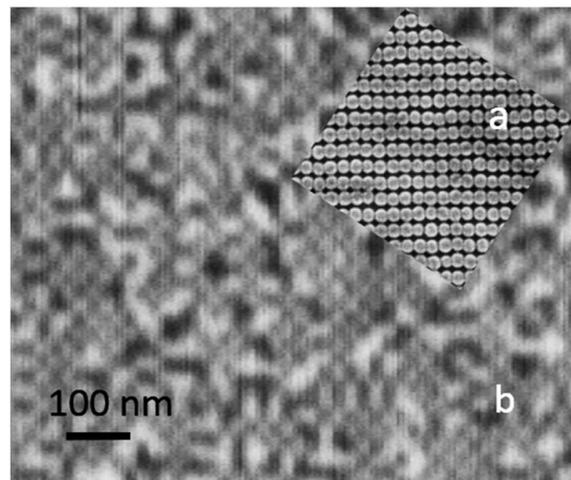

Figure 1 (a) SEM image of 1.6Td/in$^2$ BPM. (b) Typical static tester image of an AC demagnetized region.

The BPM media is prepared by AC demagnetizing the disk using an external field. A typical readback image of the ac demagnetized BPM islands acquired with the static tester is

2shown in Fig. 1b. A representative SEM image of the finished disk is shown in Fig. 1a. In the readback image we can see extended checkerboard areas where nearest neighbor dots are magnetized in opposite directions. The presence of extended checkerboard regions indicates that dipole-dipole interactions determine the resulting magnetic correlations during AC demagnetization [14], rather than switching field distributions or exchange, and indicates that the magnetic islands have good magnetic properties.

There are several advantages to using the static tester, rather than spin stand, for BPM recording experiments. The static tester can study media on small disk fragments or disks with asperities and roughness that would not be tolerated by a flying head. The static tester has excellent positioning stability with drifts that can be less than a nanometer per minute. However both of these advantages have strong caveats. Since the head is in contact with the surface, head and surface wear can occur that can degrade the head-media spacing over time. The tribological challenges are mitigated by (1) improving the surface smoothness, (2) carefully leveling the head on the disk to minimize contact forces during scanning, (3) frequently changing the head and scan location, and (4) immersing the head-disk contact area with wet lube.

The native positioning stability and repeatability of the static tester, however, is still unsuitable for BPM recording experiments at high densities. The slider that houses the read/write (RW) head is suspended on a flexible gimbal assembly. The gimbal provides an important tolerance to the head-disk alignment, but unfortunately makes the scan positioning not sufficiently linear or predictable. Variation in friction between the head and disk can slow or deviate the motion of the head by as much as 50 nm during a 30 um scan. On a given spot much of the deviation is repeatable and scan to scan line deviations of a few nm are typical.

*B. BER Measurement Scheme*

The general procedure for the recording demonstration is to evaluate the Bit Error Rate (BER) as a function of cross-track position and down-track write phase. The demonstration is a success if there is non-zero cross-track and phase margin for writing at the target BER and below. During normal HDD operation the cross track position and write phase are precisely controlled by subsystems that feedback on readback information from servo and timing patterns on the disk. The static tester experiments do not have such subsystems and instead cope with position drift and non-linearity in a different manner altogether. We have developed a scheme that works without servo feedback and where the write phase and track registration are not held fixed during the write.

Our procedure relies on being able to precisely determine the head trajectory during the write in order to reconstruct the track misregistration (TMR) and write phase for each written bit after the write. The head trajectory is determined by reading data patterns using the read head during the write. These "servo lines" acquired during the write are compared to a previously acquired "servo image" and allow an accurate reconstruction of the head-island relationship for each written bit. The details of this position extraction technique are presented in the next section.

The outline of the BER measurement procedure is as follows
1. Acquire the servo image. The servo image contains two parts: the data region (upper half of the servo image) and the servo region (bottom half of servo image)
2. Decode the servo image, i.e. determine the bit location and magnetization sign
3. Write data tracks to the data region while simultaneously reading servo lines from the servo region
4. Decode the write registration from the servo lines
5. Acquire the write region image
6. Decode the write region
7. Determine bit errors & adjacent track errors
8. Compile the BER statistics

## II. ISLAND LOCATION DETECTION

In the place of a dedicated servo system we use the BPM lattice as a positioning reference. An important part of our scheme is the ability to determine the BPM island locations from a static tester image like the one shown in Fig. 1b. In analyzing island positions it is convenient to use orthogonal basis vectors oriented along and perpendicular to the track, rather than actual lattice basis vectors. The distinction is moot for a rectangular lattice, but is important for hexagonal or skewed lattices. In this scheme the reciprocal lattice vectors are $k_{DT} = 2\pi/BP$ and $k_{CT} = 2\pi/TP$, in the down track and cross track directions respectively. The island locations are at points {x,y} wherever Eq (1) is satisfied:

$$\cos(k_{DT}x + \varphi_{DT}) + \cos(k_{CT}y + \varphi_{CT}) = 2 \quad (1)$$

For a perfect rectangular lattice the parameters $\varphi_{DT}, \varphi_{CT}, k_{DT}$, and $k_{CT}$ are constant. However, due to drifts and nonlinearities, both phases and lattice constants are spatially dependent.

In order to determine the island locations we perform a sequence of windowed Fourier transforms on the readback image. The spectral power at the lattice frequency is greatly enhanced by rectifying the readback signal by taking the absolute value of the image. We find the down-track lattice parameters by subdividing the scan line into 0.5-1 m segments and fitting the rectified readback signal to $A \cdot \cos(k_{DT}x + \varphi_{DT})$.

The fitting procedure is more flexible than finding peaks in a conventional Fast Fourier Transform (FFT) because it allows $k_{DT}$ to be resolved to a finer degree and reduces leakage that stems from the mismatch between lattice and sample points. By repeating this procedure for all segments and all scan lines we obtain how the down track lattice parameters $k_{DT}$ and $\varphi_{DT}$ vary for the whole image.

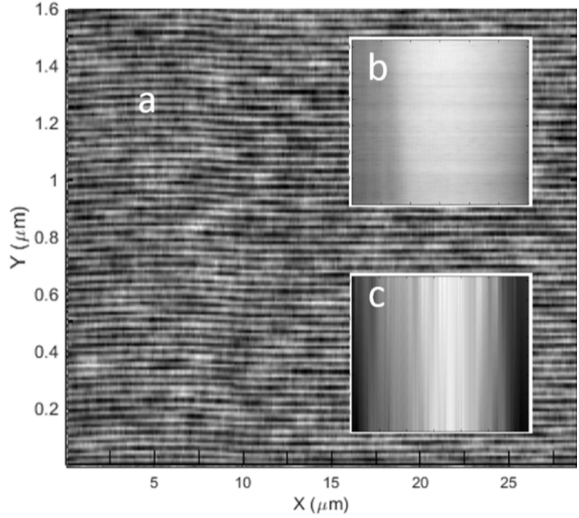

Figure 2: Island lattice analysis for the image shown in Fig. 1b (a) Absolute value of A(x,y) (see text), which is Fourier amplitude for $k_{DT}$ wave vector at specific location (b) $\Phi_{DT} = k_{DT}(x,y)x + \varphi_{DT}(x,y)$ and (c) $\Phi_{CT} = k_{CT}(x,y)x + \varphi_{CT}(x,y)$. The linear component is subtracted from $\Phi_{DT}$ and $\Phi_{CT}$. The axes range for (b) and (c) are the same as for (a) 30 µm X 1.6 µm. Images shows that dot lattice image is considerably distorted, and dot lattice should be used as a reference of detector position. Note that the aspect ratio for the images is not 1:1.

The local Fourier lattice amplitude $|A|$ and lattice phase for the down track analysis are shown in Figure 2. In the lattice amplitude image (Figure 2a) we can clearly see bright lines, corresponding to the magnetic island rows, separated by darker lines that correspond to the trenches between the island rows where the Fourier amplitude is smaller. We also see non-linearity of islands lattice in the left part of the image which is due to the non-linear motion of the head, as discussed in the previous section.

The lattice extraction in the cross-track direction is performed using the extracted down track amplitude $|A|$ as an input, rather than the readback image itself. While in principle the analysis could have been performed on the readback image, misalignment between the cross track vector and the actual lattice vector can pose complications. The result of the extraction of $\varphi_{CT}$ is shown in Figure 2c. Having extracted variation of $\varphi_{DT}, \varphi_{CT}, k_{DT}$, and $k_{CT}$ across the readback image the island locations can be determined by using the Eq. (1).

## III. BIT DECODING

If the MR reader response function is smaller than the island size, we can estimate the island magnetization using the sign of the MR reader signal at the island location. This type of threshold detector fails when the reader response function width is comparable to the island size, due to intersymbol interference (ISI) and intertrack interference (ITI) produced by the neighboring islands. If ITI or ISI is small the bit detection can be accurately handled by a 1D detection algorithms [15][16][17][18][19]. Given that each dimension of the 18 nm x 22.5 nm bit lattice is smaller than the full-width half maximum of the reader response (~30 nm x 35 nm), a 2D detection algorithm is required. For bit detection we implemented both a 2D Viterbi algorithm and a maximum likelihood decision feedback equalization type algorithm that we refer to as the 2D Sieve algorithm [20][21]. Both algorithms require the knowledge of the MR reader point response function ($R_{prf}$). First we describe the evaluation of the $R_{prf}$, followed by the description of the decoding algorithms.

### A. Reader Point Response Function

The typical approach for determining the $R_{prf}$ in 1D recording is to write a known pseudo-random bit sequence (PRBS) and fit the $R_{prf}$ to match the readback signal [22]. Similarly in 2D, the readback image $D_{img}$ can be expressed as

$$D_{img} = R_{prf} * B_{xy} \qquad (2)$$

Here the symbol '$*$' means convolution and

$$B_{xy} = \sum b_i \delta(x - x_i) \delta(y - y_i), \qquad (3)$$

$b_i$ is the bit value $\pm 1$ of island $i$, $x_i$ and $y_i$ are the island location coordinates. If the bit values are known and uncorrelated, the point response function can be calculated by taking the correlation of the readback image with the bit sequence:

$$R_{prf} \approx D_{img} \otimes B_{xy}, \qquad (4)$$

where '$\otimes$' means correlation.

Extracting the $R_{prf}$ from the 2D readback image of the Fig. 1 (b) is not possible using Eq. 4 since the bit states are not known and since the AC demagnetization anti-correlates adjacent magnetization states.

The issue of correlations can be overcome by directly deconvolving the bit states in Eq. (2) with their Fourier transforms. Deconvolution is highly sensitive to noise in the image. In order to improve the quality of the extracted $R_{prf}$ the image is first low pass filtered. Using Eq. (2),

$$\widetilde{R_{frf}} = \widetilde{R_{prf}} \cdot \widetilde{F_{LP}} = \frac{\widetilde{D_{img}} \cdot \widetilde{F_{LP}}}{\widetilde{B_{xy}}}, \qquad (5)$$

Where '$\sim$' indicates Fourier transform and $F_{LP}$ is a low-pass filter. We use $F_{LP}$ in the shape of Gauss function with a full width half maximum (FWHM) of ~ 5nm to reduce the impact of noise in the deconvolution. The $R_{frf}$ represents a filtered point spread function for the filtered image $D_{img} * F_{LP}$. Further deconvolution of $F_{LP}$ is not necessary as one can instead use the filtered image for decoding the bit values.

To solve the problem of not knowing the bit values $b_i$ a priori, the $R_{frf}$ is extracted iteratively. First we use a simple threshold detection method that estimates $b_i$ with ~90% accuracy. Then we calculate $R_{frf}$ with Eq. 5 and decode the image using the algorithms described in the next section. The process is repeated until we get satisfactory accuracy of $R_{frf}$ and of $B_{xy}$, which usually occurs within three iterations.

Figure 3a shows the $R_{frf}$ function obtained by this method. The accuracy of the $R_{frf}$ extraction and of the decoding of $b_i$ can be gauged by comparing $D_{img} * F_{LP} \otimes B_{xy}$ to $R_{frf} * B_{xy} \otimes B_{xy}$. As shown in the Figure 3b, the match between the two cross-correlations is excellent, indicating that $R_{frf}$ accuracy is unlikely to limit the decoding accuracy.

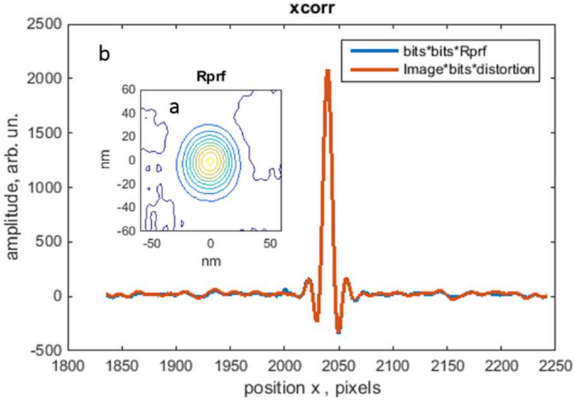

**Figure 3** (a) Contour plot for $R_{frf}$ calculated by Eq. 5. (b) Comparison of $D_{img} * F_{LP} \otimes B_{xy}$ to $R_{frf} * B_{xy} \otimes B_{xy}$, confirming the accuracy of the $R_{frf}$ extraction.

### B. 2D Viterbi algorithm

The Viterbi algorithm is a well-known algorithm for determining the most likely sequence of states that gave rise to a given set of observables [23][13]. The Viterbi algorithm gives an optimal result for systems in which there is a one-dimensional path through the sequence of observables. In the context of magnetic recording, the states are a sequence of bit values and the observables are the readback samples. The typical cost function for evaluating likelihood is the squared error between the observed readback signal and the computed response for the different Viterbi states.

In order for the Viterbi algorithm to decode optimally, the Viterbi state needs to include all of the neighboring bits that contribute to the readback signal at a given bit position. If the bit response function extends only to the two nearest neighbors, the Viterbi states will be the $2^3 = 8$ possible combinations of the three bits. The Viterbi states for neighboring sites will have two out of three bits in common. The algorithm enforces consistency by not allowing transitions between Viterbi states that would violate the bitwise sequential relationships. E.g. the predecessor of state **11**0 can only be 0**11** or 1**11** and not 001, as the corresponding bits do not match. The sequence of sites can be viewed as a chain, and the interlocking relationship between allowed Viterbi states on neighboring sites is often represented as a Trellis.

The Viterbi algorithm proceeds in two passes denoted as the forward pass and backward pass. In the forward pass the most likely predecessor state is determined for each possible successor state at each site. In the example above this means determining whether predecessor state 011 or 111 on one site has lower squared error. The squared error for the optimal predecessor state is then added to each corresponding successor state and the process repeats. As a result each proceeding decision regarding predecessors has the cost of the cascading chain of decisions already factored into it. The algorithm keeps track of the decision for what predecessor state minimizes squared error for each successor state for every site in the chain. In the backwards pass the Viterbi state for the site at the end of chain is determined by picking the one the lowest accumulated squared error. That choice enables the extraction of all of the predecessor states by using the decision table generated in the forward pass. As a result the bit sequence is determined.

In two dimensions there is no longer a simple directionality to information flow as each site needs to pass likelihood information in multiple directions [20][24]. The chain of decisions is still traversed sequentially, so sites may not necessarily have optimal information regarding the impact of decisions on all their neighbors. In addition as information is passed in two directions, care must be taken to not let duplicate information have excess weight in decisions.

We now discuss the implementation of the Viterbi algorithm we used for decoding the 1.6 Td/in2 readback data. The Viterbi state consists of nine bits: the central bit on the site and the eight nearest neighbors, labeled 1 through 9 in Figure 4a. The nine bits generate $2^9 = 512$ Viterbi states. As the bit response function had a FWHM of ~1.5-2 bits in both the cross track and down track directions, using the eight nearest neighbors captures the most of the readback response extent for the bit. The decision information is passed in two directions, as shown in Figure 4a, parallel and perpendicular to the track, which we denote as D2 and D1.

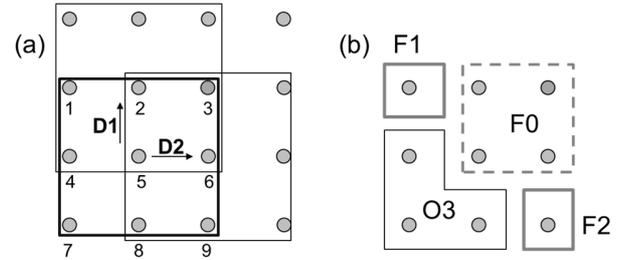

**Figure 4** (a) Schematic of the relationship between island locations and the Viterbi state for a rectangular lattice. The islands labeled 1-9 constitute the Viterbi state associated with island 5. Least squared error information is passed along directions D1 and D2 to islands 2 and 6, respectively. The boxes show the members of the Viterbi states for islands 5, 2 and 6. (b) Illustration of bits used for resolving conflicts if D1 and D2 decisions do not agree (see text).

For the site corresponding to bit 5 in Figure 4a, the two successive sites correspond to bit 2 and bit 6 in bit 5's Viterbi state for directions D1 and D2, respectively. The Viterbi states for site 2 and 5 have bits 1-6 of site 5 in common. As a result, the decision along D1 at site 5 stems from optimizing the value of bits 7-9 for each possible value of bits 1-6. We denote bits 1-6 as the decision bits, as they decide how to optimize the optimization bits 7-9. The decision for which combination of optimization bits minimizes the squared error in noted for each combination of decision bits and the respective squared error is added to the corresponding Viterbi states on site 2. An identical optimization is performed along D2, except that bits 1, 4, & 7 are the optimization bits and bits 2, 3, 5, 6, 8, & 9 are the decision bits.

The forward pass proceeds through the lattice in a linear fashion along the down track direction D2 and then wraps back around to start on the next track. At each site the squared error information for the optimized optimization bits is passed along both D1 and D2 and the subsequent optimizations factor those squared error costs into the optimizations of their respective decision bits. The chain of sites can be picked to follow a different sequence, such as starting along D1 and wrapping back or zig-zagging through the lattice. For all chains structures, care must be taken to match predecessor and successor Viterbi bits correctly.

In the backwards pass the Viterbi state for the last site with the lowest total squared error is picked in the absence of any additional information regarding the last bit in the sequence. The choice of the last Viterbi state sets off a cascade of decisions that ideally will uniquely determine all the preceding Viterbi states in the chain. However, as each site has two predecessors, one for each decision direction, it is possible that the decision along one direction conflict with the decision along the other direction.

Resolving conflicts when decisions along D1 and D2 do not match is an important part of the algorithm. Ensuring self-consistency and least-squares optimization, however, is an extremely difficult task for which there is no known exact 2D Viterbi algorithm. We employ a simple conflict resolution scheme that is illustrated in Figure 4b. We denote the Viterbi states from direction D1 and D2 as V1 and V2, respectively. Using the cumulative squared error information we pick bits F0 from whichever of V1 and V2 has lowest squared error. Next we fix bits F1 to the same value as it has in V1 and F2 to its value in V2. Having these six bits fixed we then optimize bits O3 in order to minimize least squared error with Bits F0-2 fixed. While this conflict resolution is not optimal, the procedure is quick and does not require additional passes through the chain.

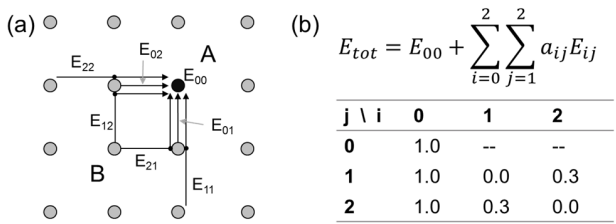

**Figure 5** (A) Schematic of sources of least squared error information for bit at location A. Seven of the sources are labeled $E_{ij}$. (B) Table showing the optimized coefficients for weighting the seven error sources when decoding the 1.6 Td/in$^2$ data.

Since squared error is passed in two directions, additional care needs to be taken to manage duplication of information. At every site the decision is based on both the squared error from the on-site comparison (direct) and from the cumulative sum of errors passed on by its two predecessor sites (indirect). The error from the predecessors can be further subdivided into its direct and indirect sources. The schematic for squared error information flow is illustrated in Figure 5a. There are seven types of squared error: three direct from the site (E00) and its two predecessors (E01 and E02) and four indirect from the predecessors of the two predecessors (E11, E22, E12, and E21). The indirect errors contain the accumulated error from past optimization. Direct error is added to the indirect error appropriately as it is passed to the successors. The seven errors are added together with weighting factors $a_{ij}$ to give a total squared error for optimizing the optimization bits,

$$E_{tot} = E_{00} + \sum_{i=0}^{2} \sum_{j=1}^{2} a_{ij} E_{ij} \qquad (6)$$

In the naïve implementations all $a_{ij}$ are set to 1. Here error from predecessors quickly multiplies as it is duplicated at each site when passed in two directions. For example, two copies of the direct error for site B in Figure 5a will be passed to site A through route E12 and E21. Sites further down the chain will have even more copies of site's B direct error.

The decoding error rate is reduced by optimizing the error coefficients $a_{ij}$. The optimization can be accomplished by minimizing the decoding error rate of simulated readback images with known bit values. For the 1.6 Td/in2 data the best fit coefficients are shown in Figure 5b. The best fit values will depend on the reader response function and will in general increase with increases in ISI and ITI.

*C. Sieve algorithm*

We now discuss a new iterative decoding algorithm which we call the Sieve algorithm. At each iteration the algorithm assigns confidence values to the bit state at each island. A bit value is assigned to each island whose confidence value exceeds a set threshold. Once a bit value is assigned the island is no longer considered in future iterations (i.e. the island is "sieved" out) and the confidence values of the islands in the local neighborhood are updated accordingly. The algorithm is repeated using the new confidence values, and with potentially updated thresholds, until all islands are assigned bit values. The Sieve algorithm has a strong similarity to decision feedback equalization [21]. The main aspect of the sieve algorithm is that decisions for bit values are made sequentially in order of highest likelihood.

In our implementation of the Sieve algorithm, the confidence value associated with an island's bit value is the magnitude of readback signal at the island center. The sign of the readback signal determines the bit value. When an island's bit value is assigned, the confidence values are updated by subtracting the readback response of the assigned island from its neighbors. To maximize accuracy the threshold is set to the maximal non-assigned island readback amplitude during each iteration. Hence only one island is assigned per iteration. If desired a threshold of 50-90% of the max readback signal could be used instead to speed up the decoding.

The assignment of bit values during the Sieve algorithm iterations is illustrated in Figure 6. At the start all islands have unassigned bit states, as indicated by white circles at island locations in Figure 6a. The inset of Figure 6a shows the histogram of the signal values at island locations. Due to ISI and ITI the histogram does not have two well-separated distributions for -1 and +1 magnetization states. The algorithm assigns bit values to the islands with largest signal magnitude and updates the readback values on the neighboring islands by subtracting off the assigned island's readback response. The resulting readback distribution represents the signal distribution if the assigned island was removed from the population and no longer contributed to ISI and ITI. Figure 6b and c shows the histogram and bit values when 25% and 50% of the islands are decoded, respectively. Blue circles show the islands that were decoded as -1 and red for +1. We see from the histograms that as the algorithm progresses the signal values for the two island magnetization states becomes increasingly better separated. The final decoded island magnetizations are shown in Figure 6d.

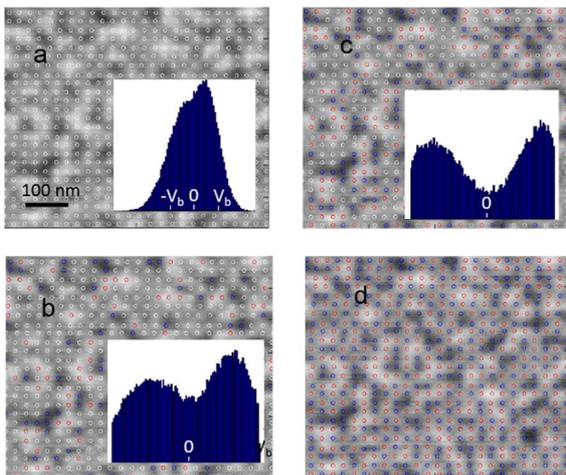

**Figure 6**: Schematic of Sieve algorithm. (a) Readback image with island locations shown as circles. White indicates the bit value has not been assigned. Inset shows the histogram of signal values at island locations. The markers +/-Vb show the signal amplitude of $R_{prf}$. (b) Island bit assignments and non-decoded island signal histogram when 25% of readback image is decoded. Red represents +1, blue -1. (c) Results after 50% of image is decoded. (d) Results after 100% of image is decoded.

In spite of its simplicity, the Sieve algorithm produces decoding results that match or exceed the performance of the 2D Viterbi algorithm implementation discussed earlier. We benchmarked the performance of the two algorithms by applying the algorithms to simulated readback images as a function of the signal to noise ratio (SNR). The noise in the simulated images comes from two sources: magnetization magnitude variability of each island, $n_M$, and readback noise, $n_R$. We refer to $n_M$ as media noise. The island magnetization values $b_i$ were randomly assigned for each run. The readback image is computed using Eqs. (7) and (8), with a $R_{prf}$ extracted from the experimental 1.6 Td/in2 readback image.

$$D_{img}(x, y) = R_{prf} * B'_{xy} + n_R(x, y) \quad (7)$$
$$B'_{xy} = \sum(b_i + n_M(i))\delta(x - x_i)\delta(y - y_i) \quad (8)$$

The readback noise and media noise are spatially uncorrelated and Gaussian distributed.

Figure 7 shows the decoding bit error rate (BER) obtained with the Sieve and 2D Viterbi algorithms as a function of media SNR. The noise and signal were calculated at the island locations. Reader noise was held fixed producing a read-to-read SNR of 22 dB. For noise dominated by media noise the Sieve algorithm outperforms the 2D Viterbi algorithm at all SNR values. For example, the Sieve algorithm can achieve a BER of 1e-2 at approximately 1 dB lower SNR than our implementation of the 2D Viterbi algorithm.

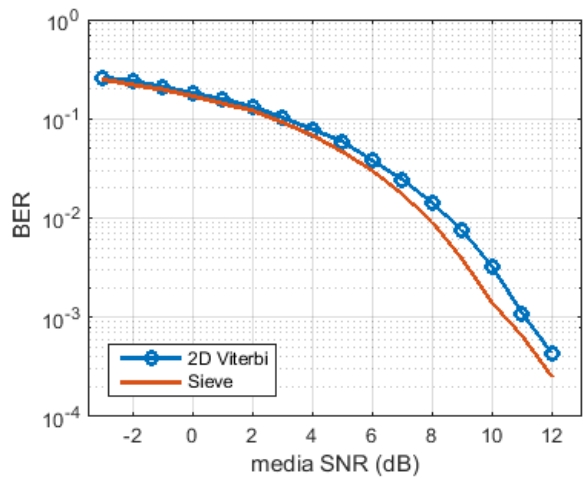

**Figure 7:** Comparison of performance 2D Viterbi vs Sieve algorithm as function of media SNR.

The performance of the 2D Viterbi algorithm implementation could be enhanced by improving how conflicts between different paths are resolved, by improving how error and conflict resolution information are propagated, and by adding equalization [16][17][18][19].. Potentially, a different implementation of 2D Viterbi could match and exceed the performance of the Sieve algorithm as implemented here. However, a big advantage of the Sieve algorithm is the simplicity of how it processes the most reliable information first.

IV. BIT ERROR RATE MEASUREMENTS

A. *Head Trajectory Detection During Writing*

Without dedicated servo features to precisely control the write head and write timing during writing, we implemented a read while write scheme to extract head trajectory information during the write process. The readback waveforms (servo lines) are compared to a reference readback servo image acquired earlier. By cross-correlating segments of the servo line to the servo image, the trajectory of the read head during the write is reconstructed. The read and write heads have a fixed spatial relationship that can be easily deduced in these measurements. From the read head trajectory, the down-track and cross-track positions of the write head is determined at each write clock edge.

Figure 8a shows a readback image that comprises both the servo image region and the neighboring data region where a data track is written. The written data track spans multiple islands in the cross track direction and is distinguished from the rest of the islands that were AC demagnetized. A segment of the servo line acquired during the writing of the track is shown as the blue curve in Figure 8b. The servo line segment is cross-correlated with the servo image and the best matched segment of the servo image is shown as the green curve in Figure 8b. The agreement is excellent and indicates that the location of the best fit segment in the servo image represents the location of the read head when the servo line segment was acquired. Figure 8a shows the trajectory of the read head as deduced by processing all the segments in the servo line.

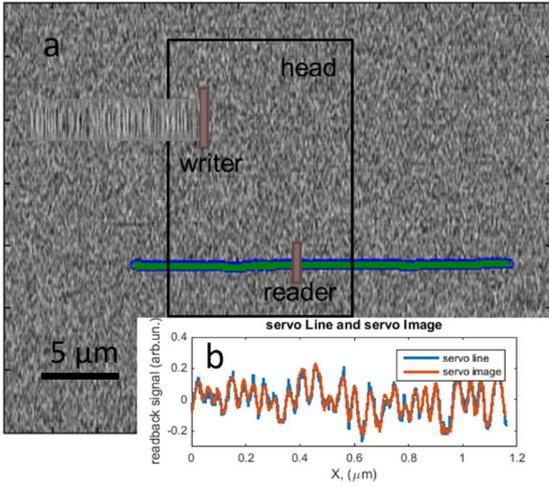

**Figure 8:** (a) Schematic of Read and Write head for track writing and Servo Line reading. (b) Comparison of a segment of Servo Line (blue) with line on Servo Image (green) demonstrating and LSD gives a very good match of signal for Servo Line segment and Servo Image.

### B. 2D Bit Error Rate Compilation

In order to assess the recording performance of the 1.6 Td/in2 bit patterned media we measure the write BER as a function of the down-track write phase and of the cross-track position of the writer relative to the BPM island. The 2D bit error rate is compiled by repeatedly writing tracks with pseudo-random bit sequences (PRBS) while monitoring the write trajectory and timing, as illustrated in Figure 8. The region with the written and neighboring islands is read back and decoded. The decoded bit values are compared to the PRBS write pattern to determine which islands were written correctly and which are in error. In addition, whether a write event caused a bit to flip is tracked by comparing bit decoding results before and after the write process. The position information acquired during the write is used to associate to each decoded bit (1) the write phase during the write attempt, (2) the relative cross-track position of the write head during the write attempt; (3) whether the write attempt flipped the bit; and (4) whether the decoded bit was written correctly. The track write process is repeated 100 times with different PRBS sequences and for different cross-track and down-track offsets. Each written track is ~1000 bits long, so typical statistics are ~$10^5$ write events with randomly distributed write registration over the whole bit cell.

The combined cross-track position, write phase, and error information are used to generate the 2D on-track error rate (OTER) map shown in Figure 9a. The map represents the probability of writing an island incorrectly for different registrations of the island relative to the write head. The blue region in Figure 9a represents the write registrations for writing individual bits at bit error rates close to the bit error rate floor of the sample (~3e-3). The shape of bit error rate floor region is curved, reflecting the curvature of the write head contours. The bit flip probability rate (BFR) is shown in Figure 9b and represents the cross-track extent of the write field of the write head. The max BFR is 50%, as expected for writing a PRBS pattern that is uncorrelated with the previous island magnetization states. Likewise, the bit error rate far away from the optimal write and cross track phase at (0,0) is 50%.

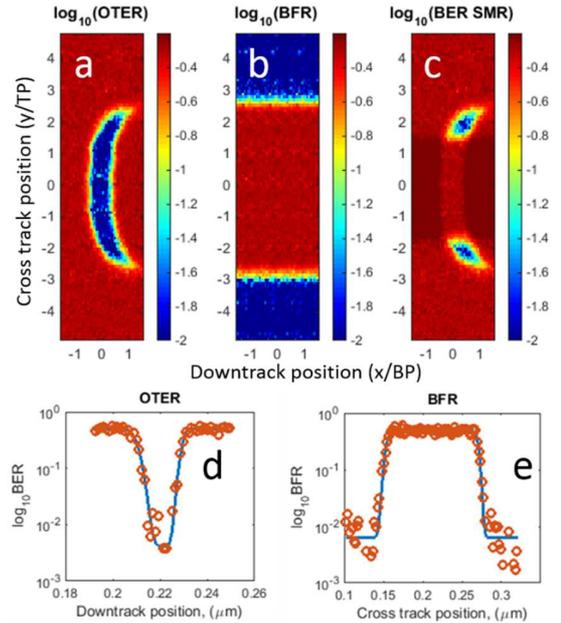

**Figure 9**: (a) On track error rate (OTER) vs. write registration (b) Bit flip probability rate (BFR) vs. write registration. (c) BER vs. write registration for shingled magnetic recording (d) OTER for zero cross track offset with fit using Eq. 8 (e) Bit flip probability rate vs. cross track position for write phase = 0 with fit as described in text.

As shown in Figure 9b, the mean write width (MWW) for the writer head is ~110 nm, spanning nearly 6 tracks. Writing adjacent tracks with low bit error rate is only possible if the tracks are written consecutively in a shingled fashion with a fixed cross-track direction. To estimate the potential of the 1.6 Td/in2 BPM for shingled magnetic recording (SMR), we add the adjacent track BFR and to the OTER to generate BER$_{SMR}$.

Figure 9c shows the BER$_{SMR}$ capability of the 1.6 Td/in2 media. Positive and negative cross-track registrations indicate shingling along increasing or decreasing track number directions, respectively. The plot shows ample margin (blue regions) for writing at close to the BER floor of 3e-3 in the shingled fashion.

### C. Recording Performance Discussion

Additional information regarding the recording performance can be extracted by fitting the OTER curve to extract effective bit position jitters using,

$$OTER = \frac{1}{4}\left(erfc\left(\frac{\delta+\frac{BP}{2}}{\sigma_{eff}}\right) + erfc\left(-\frac{\delta-\frac{BP}{2}}{\sigma_{eff}}\right)\right) \quad (9)$$

Here $BP$ is the bit pitch and $erfc$ is the complimentary error function. The effective bit position jitter $\sigma_{eff}$ combines the impact of the lithographic placement errors, $\sigma_{litho}$, the switching field distributions, $\sigma_{mag}$, the head field gradients $dH/dx$, and the position information error, $\sigma_{servo}$, on the ability to write islands correctly in quadrature (Eq. (10)):

$$\sigma_{eff}^2 = \left(\frac{\sigma_{mag}}{dH/dx}\right)^2 + \sigma_{litho}^2 + \sigma_{servo}^2 \quad (10)$$

An equivalent equation to Eq. (8) is used to fit the bit flip rate curve with $BP$ replaced by $-MWW$ and the negative and positive cross-track positions adjusted separately to the first and second term of Eq. (9), respectively. Results for fitting the cross sections of the plots in Figure 9a and Figure 9b to Eq (8) are shown in Figure 9d and Figure 9e, respectively. The extracted effective on-track and cross-track bit position jitters are $\sigma_{eff_{DT}} \sim$ 2.4 nm and $\sigma_{eff_{CT}} \sim$ 4.0 nm, respectively.

These effective jitters are dominated by the magnetic switching field distribution component of Eq. (9). We estimate that the effective write field gradients are 300 Oe/nm down-track and 150 Oe/nm cross-track and the island switching field distribution is measured to be 540 Oe. We obtain $\sigma_{mag_{DT}} \sim =$ 1.8 nm and $\sigma_{mag_{CT}} =$ 3.6 nm. The position distributions are estimated from the analysis of scanning electron microscopy images: $\sigma_{litho_{DT}} =$ 1.1 nm and $\sigma_{litho_{CT}} =$ 1.2 nm. This leads to $\sigma_{servo_{DT}} \sim$ 1 nm and $\sigma_{servo_{CT}} \sim$ 1.2 nm, which reflects the accuracy of the write position detection scheme used in this experiment.

### D. Readback and Media Noise

In this section, we describe the estimation of readback noise and island-to-island magnetization variation (media noise) from the 1.6 Td/in2 readback data. To estimate the readback signal-to-noise ratio (SNRrd), we acquired 20 consecutive readback images of the same 30 um^2 spot in which the islands were prepared by writing PRBS data in a shingled fashion. To remove the image to image drift we subdivided each image into 100 fragments and aligned them separately in consecutive images. For each fragment we averaged the aligned images. Reader noise was defined as root mean square (RMS) deviation from the averaged image. Signal is defined as the standard deviation of the averaged image. The resulting $SNR_{rd}$ extracted for the 1.6 Td/in2 sample is 16 dB. The readback noise includes not only reader noise but also contribution from non-repeatable variation of head motion. Reducing the mismatch between the reader width and the medium track pitch is expected to improve $SNR_{rd}$.

The media noise was evaluated by comparing the difference between readback images and simulated readback images generated from the best extracted $R_{prf}$. The signal is defined as the standard deviation of the simulated image. The noise is defined as the standard deviation of the difference between the readback and simulated image. The extracted $SNR_{tot}$ is 8.6 dB and includes media noise and read noise. Assuming that read and media noise add in quadrature we find that $SNR_{media}=$ 9.5 dB. Comparing to the simulated channel performance as a function of BER (Figure 7), the decoding BER for an $SNR_{media}$ of 9.5 dB is ~ 2e-3, which is close to the measured BER floor.

### V. CONCLUSION

In this paper we present several analytic and experimental techniques that enabled recording studies of 1.6 Td/in2 BPM. We described a method for determining island locations in readback images as well as technique for determining the write head trajectory during a track write. We compared two different 2D bit decoding techniques. The resulting study demonstrated the feasibility of recording to at 1.6 Td/in2 with a 2D BER below 1e-2 using shingled magnetic recording.